\documentclass[review]{elsarticle}
\usepackage{amsmath}
\usepackage{graphicx}
\usepackage[unicode,bookmarks,bookmarksopen,bookmarksopenlevel=2,colorlinks,linkcolor=blue,citecolor=green]{hyperref}

\newtheorem{example}{Example}

\begin{document}
\begin{frontmatter}
\title{A new class of solutions for the multi-component extended Harry
Dym equation\footnote{Dedicated to the 80th birthday of A.B. Shabat}
}
\author[MMa]{Michal Marvan}
\author[MPa,MPb,MPc]{Maxim V. Pavlov}
\address[MMa]{Mathematical Institute in Opava, Silesian university in Opava,
  Na Rybn\'{\i}\v{c}ku 1, 746 01 Opava, Czech Republic}
\address[MPa]{Novosibirsk State University, Pirogova street 2, 630090, Novosibirsk, Russia}
\address[MPb]{Sector of Mathematical Physics, Lebedev Physical Institute of Russian Academy of Sciences,
  Leninskij Prospekt 53, 119991 Moscow, Russia}
\address[MPc]{Department of Applied Mathematics, National Research Nuclear University MEPHI,
  Kashirskoe Shosse 31, 115409 Moscow, Russia}

\begin{abstract}
We construct a point transformation between two integrable systems, the multi-component Harry Dym 
equation and the multi-component extended Harry Dym equation, that does not preserve the class 
of multi-phase solutions. 
As a consequence we obtain a new type of wave-like solutions,
generalising the~multi-phase solutions of the multi-component extended Harry Dym equation. 
Our construction is easily transferable to other integrable systems with analogous properties.
\end{abstract}

\end{frontmatter}

\section{Introduction}

\label{sec-intro}

In a number of papers \cite{AF1,AF2,AF3}, integrable systems associated with
the energy-dependent linear Schr\"{o}dinger equation%
\begin{equation*}
\biggl(\sum_{m=0}^{M}\epsilon _{m}\lambda ^{m}\biggr)\psi _{xx}=\biggl(%
\sum_{m=0}^{M}v_{m}\lambda ^{m}\biggr)\psi 
\end{equation*}%
were investigated in details. In these papers, two main classes were
selected by the conditions $\epsilon _{M}=0$, $v_{M}=1$ (the so called
\textquotedblleft multi-component KdV systems\textquotedblright ) and $%
\epsilon _{M}=0$, $v_{0}=1$ (the so called \textquotedblleft multi-component
extended Harry Dym systems\textquotedblright\ or \textquotedblleft
multi-component Camassa--Holm systems\textquotedblright ). In the present
paper we consider another class determined by a sole restriction $v_{M}=0$
(the so called \textquotedblleft multi-component Harry Dym
systems\textquotedblright ~\eqref{kappalong} or \textquotedblleft
multi-component Hunter--Saxton equations\textquotedblright ). We show that
the multi-component Harry Dym equations are connected with the
multi-component extended Harry Dym equations by the $\kappa $-transformation
introduced below, see~\eqref{kappatr}. The method presented here is
applicable to any solutions. Without loss of generality and for simplicity
we restrict our consideration to multi-phase solutions only. Applying the $%
\kappa $-transformation to the multi-phase solutions of multi-component
Harry Dym systems we obtain a new class of solutions of multi-component
extended Harry Dym systems, which we call the $\kappa $-deformed multi-phase
solutions. This new class of solutions cannot be obtained as a reduction of
multi-phase solutions. Whilst multi-phase solutions belong to the so called
\textquotedblleft solitonic sector\textquotedblright\ (i.e., the case of
reflectionless potentials), the new class of solutions (the $\kappa $%
-deformed multi-phase solutions) has a rapidly increasing behaviour with
respect to $x$.

\begin{example} \rm The extended Harry Dym equation ($\kappa$ is an arbitrary
constant, $\kappa \neq 0$)%
\begin{equation*}
v_{t}=-\tfrac{1}{2}(\partial _{x}^{2}-\kappa ^{2})( v^{-1/2}) _{x}
\end{equation*}%
possesses the one-phase solution ($c$ is an arbitrary constant)%
\begin{equation*}
v=\frac{1}{c^{2}r^{2}},
\end{equation*}%
where $r$ is a function of $\theta =x+ct$ determined implicitly by ($%
s_{0},s_{1}$ are arbitrary constants)%
\begin{equation*}
\theta =\overset{r}{\int }\frac{\sqrt{\lambda }\,d\lambda }{\sqrt{%
4c^{-2}+s_{1}\lambda +s_{0}\lambda ^{2}+\kappa ^{2}\lambda ^{3}}};
\end{equation*}%
and simultaneously the $\kappa $-deformed one-phase solution%
\begin{equation*}
v=\frac{e^{2\kappa x}}{c^{2}R^{2}},
\end{equation*}%
where the function $R(\vartheta )$ is determined implicitly by%
\begin{equation*}
\frac{e^{\kappa x}-1}{\kappa }+ct=\vartheta =\overset{R}{\int }\frac{\sqrt{%
\lambda }\,d\lambda }{\sqrt{4c^{-2}+s_{1}\lambda +s_{0}\lambda ^{2}}}.
\end{equation*}
These solutions are significantly different: the first solution is
essentially one-dimensional (see Figure~\ref{fig:sd-kappa=7}) and is
obtained by a regular procedure (one can look for a travelling wave
reduction determined by the ansatz $v(\theta )$, where $\theta =x+ct$),
while the second solution is two-dimensional (see Figure~\ref%
{fig:v-kappa=005}) and is obtained (see below) by means of an invertible
point transformation between the extended Harry Dym equation and its
high-frequency limit ($\kappa =0$), which is the well-known Harry Dym
equation.

\begin{figure}[h]
\begin{center}
\includegraphics[scale=0.5]{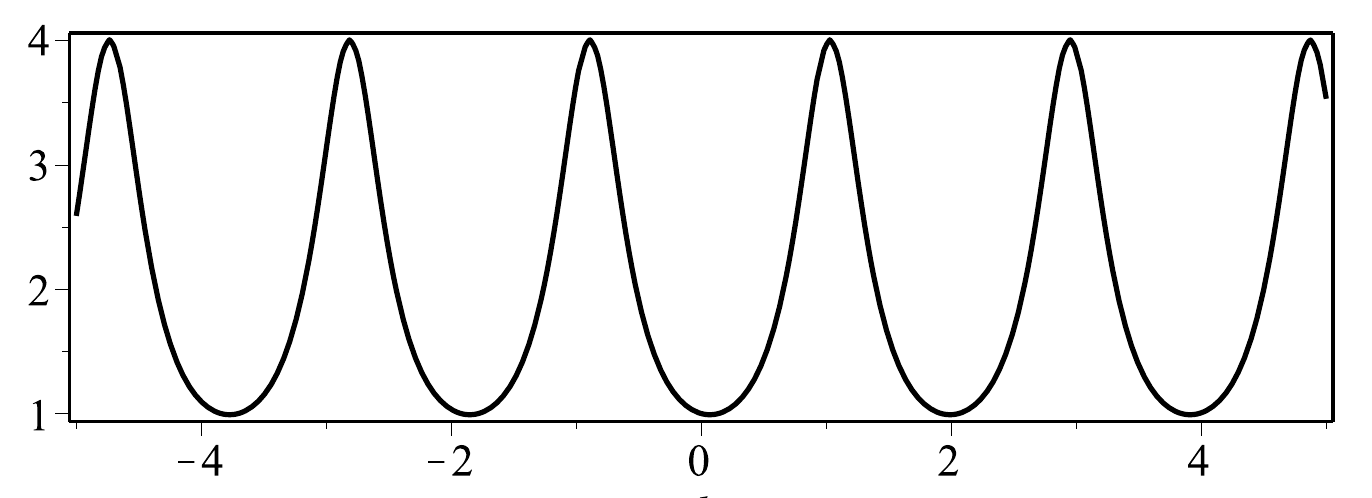} \\[0pt]
\end{center}
\caption{The unit-speed ($c = 1$) travelling wave $v(\protect\theta)$ for $%
s_1 = 12$, $s_0 = 8$, $\protect\kappa = 0.14$}
\label{fig:sd-kappa=7}
\end{figure}

\begin{figure}[tbp]
\begin{center}
\includegraphics[scale=0.35]{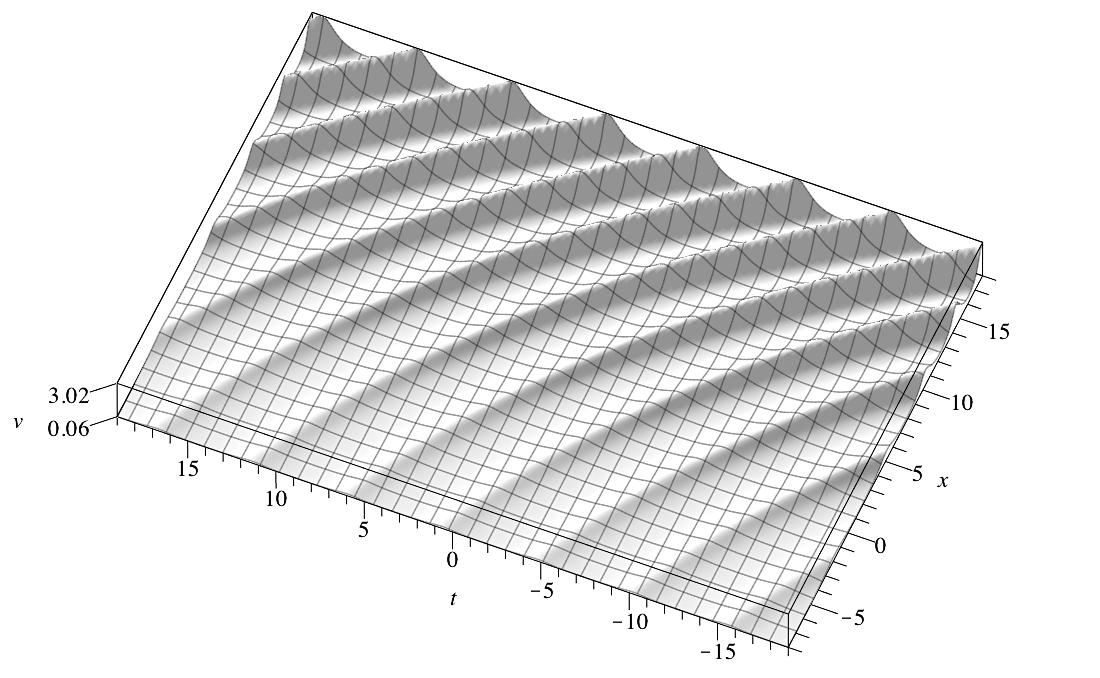} \\[0pt]
\end{center}
\caption{The $\protect\kappa$-deformed one-phase solution $v(t,x)$
for $c = \protect\sqrt{2}$, $s_1 = 3$, $s_0 = 1$, $\protect\kappa = 0.05$}
\label{fig:v-kappa=005}
\end{figure}

Many integrable systems have such an invertible point transformation that
connects them with their high-frequency limits. So, once a one-phase (or a
multi-phase) solution is found, one can construct the so-called $\kappa $%
-deformed solution following our approach.

Moreover, in the particular case%
\begin{equation*}
4c^{-2}+s_{1}\lambda +s_{0}\lambda ^{2}=s_{0}(\lambda -\lambda _{1})^{2},
\end{equation*}%
the second solution can be found in elementary functions, i.e.%
\begin{equation*}
\frac{e^{\kappa x}-1}{\kappa }+ct=\vartheta =\frac{1}{\sqrt{s_{0}}}\overset{R%
}{\int }\frac{\sqrt{\lambda }\,d\lambda }{\lambda -\lambda _{1}}=\frac{1}{%
\sqrt{s_{0}}}\biggl(2\sqrt{R}-2\sqrt{\smash[b]{\lambda_1}}\mathop{\rm
arctanh}\sqrt{\frac{R}{\smash[b]{\lambda_1}}}\biggr).
\end{equation*}

In another particular case%
\begin{equation*}
4c^{-2}+s_{1}\lambda +s_{0}\lambda ^{2}+\kappa ^{2}\lambda ^{3}=\kappa
^{2}(\lambda -\lambda _{2})(\lambda -\lambda _{3})^{2},
\end{equation*}%
the first solution can be found in elementary functions, i.e.%
\begin{eqnarray*}
\theta  &=&\frac{1}{\kappa }\overset{r}{\int }\frac{\sqrt{\lambda }%
\,d\lambda }{(\lambda -\lambda _{3})\sqrt{\lambda -\lambda _{2}}} \\
&=&\frac{1}{\kappa }\biggl(\ln (r-\tfrac{1}{2}\lambda _{2}+\sqrt{r\mathstrut 
}\sqrt{{r-\lambda _{2}}}) \\
&&-\frac{\sqrt{\lambda _{3}}}{\sqrt{\lambda _{3}-\lambda _{2}}}\ln \left( 
\frac{2r\lambda _{3}-\lambda _{2}(r+\lambda _{3})+2\sqrt{\lambda _{3}}\sqrt{%
\lambda _{3}-\lambda _{2}}\sqrt{r}\sqrt{r-\lambda _{2}}}{r-\lambda _{3}}%
\right) \biggr).
\end{eqnarray*}
Thus, both types of solutions have completely different behaviour.
\end{example}

In this paper we restrict our consideration to the simplest integrable
system whose multi-phase solutions are associated with hyperelliptic Riemann
surfaces: the multi-component extended Harry Dym equation.


\section{A Special Class of Integrable Systems}

\label{sec:integrability}

The energy dependent linear Schr\"{o}dinger equation%
\begin{equation}
\psi _{xx}=U\psi  \label{eden}
\end{equation}%
was recently \cite{energy} investigated for the special case (with
respect to the spectral parameter~$\lambda $)%
\begin{equation}
U(x,t,\lambda )=\sigma +\frac{u^{1}}{\lambda }+\frac{u^{2}}{\lambda ^{2}}+%
\frac{u^{3}}{\lambda ^{3}}+...,  \label{one}
\end{equation}%
where $\sigma $ is an arbitrary constant. If $\sigma \neq 0$, then the 
parameter $\sigma $ can be fixed to 1 by an appropriate scaling of 
independent variables and dependent functions without loss of generality.

Integrable systems associated with (\ref{eden}) can be obtained from the
compatibility condition $(\psi _{t})_{xx}=(\psi _{xx})_{t}$, where%
\begin{equation}
\psi _{t}=a\psi _{x}-\tfrac{1}{2}a_{x}\psi .  \label{psit}
\end{equation}%
The compatibility condition $(\psi _{t})_{xx}=(\psi _{xx})_{t}$ yields the
relationship%
\begin{equation*}
U_{t}=\left( -\tfrac{1}{2}\partial _{x}^{3}+2U\partial _{x}+U_{x}\right) a,
\end{equation*}%
which leads to the dispersive integrable chain ($\xi $ is an integration
constant)%
\begin{equation}
u_{t}^{k}=u_{x}^{k+1}+a^{1}u_{x}^{k}+2u^{k}a_{x}^{1},\text{ }k=1,2,...,\quad
u^{1}+\xi =\tfrac{1}{2}a_{1,xx}-2\sigma a_{1},  \label{genCH}
\end{equation}%
where $a=\lambda +a^{1}$ and the function $U$ is determined by (\ref{one}).
This dispersive integrable chain can be reduced to $M$-component integrable
dispersive systems by simple reductions $u^{M+1}=0$ for any natural number $%
M $. This means that one should consider the linear problem (\ref{eden}), (%
\ref{psit}), where%
\begin{equation}
U(x,t,\lambda )=\sigma +\frac{u^{1}}{\lambda }+\frac{u^{2}}{\lambda ^{2}}%
+...+\frac{u^{M}}{\lambda ^{M}}  \label{dva}
\end{equation}%
instead of (\ref{one}). If $M=1$, one obtains the remarkable Camassa--Holm
equation%
\begin{equation*}
u_{t}^{1}=a_{1}u_{x}^{1}+2u^{1}a_{1,x},\quad u^{1}+\xi =\tfrac{1}{2}%
a_{1,xx}-2\sigma a_{1},
\end{equation*}%
if $M>1$, the multi-component generalisation of the Camassa--Holm equation is%
\begin{equation}
u_{t}^{k}=u_{x}^{k+1}+a^{1}u_{x}^{k}+2u^{k}a_{x}^{1},\text{ }%
k=1,2,...,M-1,\quad u_{t}^{M}=a^{1}u_{x}^{M}+2u^{M}a_{x}^{1},
\label{multiCH}
\end{equation}%
where again $u^{1}+\xi =\frac{1}{2}a_{1,xx}-2\sigma a_{1}$.

Below we also investigate the special case $\sigma =0$ and discuss the
relationship between integrable systems determined by both choices $\sigma =0
$ and $\sigma \neq 0$. The corresponding dispersive integrable chain reduces
to the form (cf. (\ref{genCH}))%
\begin{equation}
u_{t}^{k}=u_{x}^{k+1}+a^{1}u_{x}^{k}+2u^{k}a_{x}^{1},\text{ }k=1,2,...,\text{
}u^{1}+\xi =\tfrac{1}{2}a_{1,xx}.  \label{genHS}
\end{equation}%
So, the main difference between (\ref{genCH}) and (\ref{genHS}) is a
difference between the constraints $u^{1}+\xi =\frac{1}{2}a_{1,xx}-2\sigma
a_{1}$ and $u^{1}+\xi =\tfrac{1}{2}a_{1,xx}$. Again if $M=1$, one can obtain
the Hunter--Saxton equation%
\begin{equation*}
u_{t}^{1}=a_{1}u_{x}^{1}+2u^{1}a_{1,x},\quad u^{1}+\xi =\tfrac{1}{2}a_{1,xx},
\end{equation*}%
which is a high frequency limit of the Camassa--Holm equation (see detail in 
\cite{Dai}). If $M>1$, the multi-component generalisation of the
Hunter--Saxton equation is (cf. (\ref{multiCH}))%
\begin{equation*}
\begin{aligned} &u_{t}^{k}=u_{x}^{k+1}+a^{1}u_{x}^{k}+2u^{k}a_{x}^{1},\quad
k=1,2,...,M-1,\\ &u_{t}^{M}=a^{1}u_{x}^{M}+2u^{M}a_{x}^{1}, \end{aligned}
\end{equation*}%
where again $u^{1}+\xi =\frac{1}{2}a_{1,xx}$. If instead of the choice $%
a=\lambda +a_{1}$ we consider the dependence $a=a_{-1}/\lambda $, then%
\begin{equation}
\begin{aligned} &u_{t}^{1}=-\tfrac{1}{2}(a_{-1})_{xxx}, \\
&u_{t}^{k}=a_{-1}u_{x}^{k-1}+2u^{k-1}(a_{-1})_{x},\quad k=2,\dots ,M, \\
&a_{-1}=(u^{M})^{-1/2}. \end{aligned}  \label{hay}
\end{equation}%
If $M=1$, then this is well-known Harry Dym equation; if $M>1$, then this
system~\eqref{hay} will be called the multi-component Harry Dym equation.

Below we show that integrable systems (\ref{hay}) associated with the energy
dependent linear Schr\"{o}dinger equation (see (\ref{eden}) and (\ref{dva})
in the limit $\sigma =0$)%
\begin{equation}
\psi _{xx}=\left( \frac{u^{1}}{\lambda }+\frac{u^{2}}{\lambda ^{2}}+...+%
\frac{u^{M}}{\lambda ^{M}}\right) \psi  \label{zero}
\end{equation}%
can be interpreted as a high frequency limit of the so called
multi-component extended Harry Dym equation (see detail below).

Indeed, one can consider the linear spectral problem (\ref{zero}) written in
the form%
\begin{equation}
\psi _{zz}=\left( \frac{u^{1}}{\lambda }+\frac{u^{2}}{\lambda ^{2}}+...+%
\frac{u^{M}}{\lambda ^{M}}\right) \psi .  \label{dzero}
\end{equation}%
Then we apply the point transformation%
\begin{equation}
z=\frac{e^{\kappa x}-1}{\kappa },  \label{zet}
\end{equation}%
where $\kappa $ is an arbitrary parameter. Then $\partial _{z}\rightarrow
e^{-\kappa x}\partial _{x}$. If $\kappa \rightarrow 0$, then $z(x,\kappa
)\rightarrow x$.

Under transformation (\ref{zet}) the linear spectral problem (\ref{dzero})
becomes (see (\ref{eden}) and (\ref{one}))%
\begin{equation}
\varphi _{xx}=\left( \sigma +\frac{v^{1}}{\lambda }+\frac{v^{2}}{\lambda ^{2}%
}+...+\frac{v^{M}}{\lambda ^{M}}\right) \varphi ,  \label{kapa}
\end{equation}%
where $\sigma =\kappa ^{2}/4$, $v^{k}=u^{k}e^{2\kappa x}$ and $\psi =\varphi
\exp (\kappa x/2)$. The high frequency limit $\kappa \rightarrow 0$ reduces
the above linear problem to (\ref{zero}). We illustrate this property for
the multi-component extended Harry Dym equation%
\begin{equation}
\begin{aligned} &v_{t}^{1}=-\tfrac{1}{2}(\partial _{x}^{3}-\kappa
^{2}\partial _{x})\tilde{a}_{-1}, \\
&v_{t}^{k}=\tilde{a}_{-1}v_{x}^{k-1}+2v^{k-1}(\tilde{a}_{-1})_{x},\quad
k=2,\dots,M, \\ &\tilde{a}_{-1}=(v^{M})^{-1/2}. \end{aligned}
\label{kappalong}
\end{equation}%
This system follows from the compatibility condition $(\varphi
_{t})_{xx}=(\varphi _{xx})_{t}$, where the function $\varphi $ is a common
solution of two linear equations, i.e. (\ref{kapa}) and (cf. (\ref{psit}))%
\begin{equation*}
\varphi _{t}=\frac{1}{\lambda }\left( \tilde{a}_{-1}\varphi _{x}-\frac{1}{2}(%
\tilde{a}_{-1})_{x}\varphi \right) .
\end{equation*}%
The high frequency limit $\kappa \rightarrow 0$ leads to the system (\ref%
{hay}). Now we apply point transformation (\ref{zet}) to system (\ref{hay})
written in the form (here we simply replaced $x$ by $z$)%
\begin{equation}
\begin{aligned} &u_{t}^{1}=-\tfrac{1}{2}(a_{-1})_{zzz},
\\&u_{t}^{k}=a_{-1}u_{z}^{k-1}+2u^{k-1}(a_{-1})_{z},\quad k=2,\dots ,M, \\
&a_{-1}=(u^{M})^{-1/2} \end{aligned}  \label{haz}
\end{equation}%
Then we again obtain system (\ref{kappalong}), where $a_{-1}=\tilde{a}%
_{-1}e^{\kappa x},$ $v^{k}=u^{k}e^{2\kappa x}$. Thus integrable systems (\ref%
{kappalong}) and (\ref{haz}) are connected with each other by the point
transformation 
\begin{equation}
z=\frac{e^{\kappa x}-1}{\kappa }, \quad u^k = e^{-2\kappa x} v^{k}
\label{kappatr}
\end{equation}%
and simultaneously system (\ref{hay}) $\equiv$ (\ref{haz}) is a high
frequency limit of system (\ref{kappalong}).

\section{Multi-Phase Solutions and a High Frequency Limit}

\label{sec:high}

To illustrate a difference between the general case $\kappa \neq 0$ and its
high-frequency limit $\kappa =0$, in this section we consider multi-gap
solutions of the multi-component extended Harry Dym equation

\begin{equation}
v_{t}^{1}=-\tfrac{1}{2}(\partial _{x}^{3}-\kappa ^{2}\partial _{x})\tilde{a}%
_{-1},\quad v_{t}^{k}=\tilde{a}_{-1}v_{x}^{k-1}+2v^{k-1}(\tilde{a}%
_{-1})_{x},\text{ }k=2,...,M,  \label{ehd}
\end{equation}%
where $\tilde{a}_{-1}=(v^{M})^{-1/2}$.

In this case linear problem (\ref{eden}), (\ref{psit}) reduces to the form%
\begin{equation}
\lambda ^{M}(2\phi \phi _{xx}-\phi _{x}^{2})=\Bigl( \kappa ^{2}\lambda ^{M}+4%
\overset{M}{\underset{m=1}{\sum }}v_{m}\lambda ^{M-m}\Bigr) \phi
^{2}-S(\lambda ),\quad \phi _{t}=a\phi _{x}-a_{x}\phi ,  \label{i}
\end{equation}%
where $\phi =\psi \psi ^{+}$ (here $\psi $ and $\psi ^{+}$ are two linearly
independent solutions), $a=\tilde{a}_{-1}/\lambda $ and $S(\lambda )$ is a
polynomial expression with constant coefficients.

As usual, finite-gap solutions connected with hyperelliptic Riemann surfaces
can be constructed in several steps:

\textbf{1}. We seek polynomial solutions (with respect to the spectral
parameter $\lambda $) for the function $\phi $ in the factorised form%
\begin{equation}
\phi =\overset{N}{\underset{m=1}{\prod }}(\lambda -r^{m}(x,t)),  \label{b}
\end{equation}%
where $N$ is an arbitrary natural number.

\textbf{2}. Since function $\phi $ is a polynomial of the degree$\ N$, the
dependence $S(\lambda )$ is a polynomial of the degree $2N+M$, i.e.%
\begin{equation}
S(\lambda )=s_{2N+M-1}+s_{2N+M-2}\lambda +\cdots +s_{1}\lambda
^{2N+M-2}+s_{0}\lambda ^{2N+M-1}+s_{-1}\lambda ^{2N+M},  \label{r}
\end{equation}%
where $s_{-1}=\kappa ^{2},$ while other $s_{k}$ are \textquotedblleft
integration constants\textquotedblright .

\textbf{3}. Expanding $\phi $ by virtue of (\ref{b}) in the first equation
of (\ref{i}) with respect to the spectral parameter $\lambda $, one can find
expressions for field variables $v_{k}$. Indeed, substituting (\ref{b}) into
(\ref{i}), one obtains%
\begin{equation}
v^{k}=\frac{1}{4}\sum_{m=0}^{M-k}\frac{s_{2N+k-1+m}B_{m}}{\left(
\prod_{n=1}^{N}r^{n}\right) ^{m+2}},\quad k=1,...,M,  \label{vi}
\end{equation}%
where%
\begin{equation}
B_{0}=1,\quad B_{k}= \underset{k_{1} + \cdots+ k_{N} = k}{\underset{%
k_{1}\geq 0, \dots , k_{N}\geq 0}{\sum }}%
\prod_{m=1}^{N}(k_{m}+1)(r^{m})^{k-k_{m}}.  \label{bn}
\end{equation}

\textbf{4}. Following B.A. Dubrovin \cite{dubr1,dubr2}, we consider the limit $%
\lambda \rightarrow r^{i}(x,t)$ of the two equations (\ref{i}). This
straightforward computation yields two autonomous systems%
\begin{equation}
r_{x}^{i}=\frac{1}{\prod_{m\neq i}(r^{i}-r^{m})}\sqrt{\frac{S(r^{i})}{{%
(r^{i})^{M}}}},\quad r_{t}^{i}=\frac{a^{i}(\mathbf{r})}{\prod_{m\neq
i}(r^{i}-r^{m})}\sqrt{\frac{S(r^{i})}{(r^{i})^{M}}},  \label{e}
\end{equation}%
where $a^{i}(\mathbf{r})=a(\lambda ,\mathbf{r})|_{\lambda =r^{i}}$. In our
case (see (\ref{vi}), $k=M$)%
\begin{equation*}
a^{i}(\lambda ,\mathbf{r})=\frac{\tilde{a}_{-1}}{r^{i}}=\frac{(v^{M})^{-1/2}%
}{r^{i}}=\frac{2\prod_{m=1}^{N}r^{m}}{r^{i}\sqrt{s_{2N+M-1}}}.
\end{equation*}

\textbf{5}. A straightforward integration of (\ref{e}) implies multi-phase
solutions of (\ref{kappalong}) written in the implicit form\footnote{%
Explicit formulae for more wide class of integrable systems, whose
multi-phase solutions are associated with hyperelliptic Riemann surfaces,
were obtained in \cite{Alber}}%
\begin{equation}
x=\underset{m=1}{\overset{N}{\sum }}\,\overset{r^{m}}{\int }\frac{\lambda
^{M/2+N-1}d\lambda }{\sqrt{S(\lambda )}},\quad t=\frac{\sqrt{s_{2N+M-1}}%
}{2}\underset{m=1}{\overset{N}{\sum }}\,\overset{r^{m}}{\int }\frac{\lambda
^{M/2}d\lambda }{\sqrt{S(\lambda )}},  \label{xt}
\end{equation}%
\begin{equation*}
0=\underset{m=1}{\overset{N}{\sum }}\,\overset{r^{m}}{\int }\frac{\lambda
^{M/2+k}d\lambda }{\sqrt{S(\lambda )}},\quad k=1,\ldots ,N-2.
\end{equation*}

\textbf{Remark}: If $N=1$, then a one-phase solution is parameterised by a
single function $r(\theta )$. Namely,%
\begin{equation*}
v_{k}=\frac{1}{4}\sum_{m=1}^{M-k+1}ms_{m+k}r^{-m-1},
\end{equation*}%
where the function $r(\theta )$ is determined by the relationship (here $%
s_{M+1}=4c^{-2}$)%
\begin{equation*}
x+ct=\theta =\overset{r}{\int }\frac{\lambda ^{M/2}\,d\lambda }{\sqrt{%
s_{M+1}+s_{M}\lambda +\dots +s_{1}\lambda ^{M}+s_{0}\lambda ^{M+1}+\kappa
^{2}\lambda ^{M+2}}}.
\end{equation*}

\subsection{Finite-Gap Solutions and $\protect\kappa $-Transformation}

Now we can compare finite-gap solutions for both cases $\kappa =0$ and $%
\kappa \neq 0$. So in the case $\kappa =0$ we have for system (\ref{haz})
multi-phase solutions (cf. (\ref{vi}) and (\ref{bn}))%
\begin{equation*}
u^{k}=\frac{1}{4}\sum_{m=0}^{M-k}\frac{s_{2N+k-1+m}\tilde{B}_{m}}{\left(
\prod_{n=1}^{N}R^{n}\right) ^{m+2}},\quad k=1,...,M,
\end{equation*}%
where%
\begin{equation*}
\tilde{B}_{0}=1,\quad \tilde{B}_{k}= \underset{k_{1} + \cdots+ k_{N} = k}{%
\underset{k_{1}\geq 0, \dots , k_{N}\geq 0}{\sum }}
\prod_{m=1}^{N}(k_{m}+1)(R^{m})^{k-k_{m}}.
\end{equation*}%
The dependencies $R^{k}(z,t)$ are presented in implicit form (cf. (\ref{xt}))%
\begin{equation}
z=\underset{m=1}{\overset{N}{\sum }}\,\overset{R^{m}}{\int }\frac{\lambda
^{M/2+N-1}d\lambda }{\sqrt{P(\lambda )}},\quad 
t=\frac{\sqrt{s_{2N+M-1}}%
}{2}\underset{m=1}{\overset{N}{\sum }}\,\overset{R^{m}}{\int }\frac{\lambda
^{M/2}d\lambda }{\sqrt{P(\lambda )}},  \label{zt}
\end{equation}%
\begin{equation*}
0=\underset{m=1}{\overset{N}{\sum }}\,\overset{R^{m}}{\int }\frac{\lambda
^{M/2+k}d\lambda }{\sqrt{P(\lambda )}},\quad k=1,\ldots ,N-2,
\end{equation*}%
where (cf. (\ref{r}))%
\begin{equation*}
P(\lambda )=s_{2N+M-1}+s_{2N+M-2}\lambda +\cdots +s_{1}\lambda
^{2N+M-2}+s_{0}\lambda ^{2N+M-1}.
\end{equation*}

Under transformation (\ref{zet}) system (\ref{haz}) becomes (\ref{kappalong}%
), while multi-phase solutions of system (\ref{haz}) take the form (we
remind the reader that $v^{k}=u^{k}e^{2\kappa x}$)%
\begin{equation*}
v^{k}=\frac{1}{4}e^{2\kappa x}\sum_{m=0}^{M-k}\frac{s_{2N+k-1+m}\tilde{B}_{m}%
}{\left( \prod_{n=1}^{N}R^{n}\right) ^{m+2}},\quad k=1,...,M,
\end{equation*}%
where the dependencies $R^{k}(x,t)$ are presented in implicit form (cf. (\ref%
{xt}), (\ref{zt}))%
\begin{equation*}
\frac{e^{\kappa x}-1}{\kappa }=\underset{m=1}{\overset{N}{\sum }}\,\overset{%
R^{m}}{\int }\frac{\lambda ^{M/2+N-1}d\lambda }{\sqrt{P(\lambda )}},\quad
t=\frac{\sqrt{s_{2N+M-1}}}{2}\underset{m=1}{\overset{N}{\sum }}\,\overset{%
R^{m}}{\int }\frac{\lambda ^{M/2}d\lambda }{\sqrt{P(\lambda )}},
\end{equation*}%
\begin{equation*}
0=\underset{m=1}{\overset{N}{\sum }}\,\overset{R^{m}}{\int }\frac{\lambda
^{M/2+k}d\lambda }{\sqrt{P(\lambda )}},\quad k=1,\ldots ,N-2.
\end{equation*}

Thus, we found a new type of solutions of multi-component Harry Dym equation
(\ref{kappalong}), which do not coincide with the corresponding multi-phase
solutions (\ref{xt}).

\subsection{The One-Phase Solution}

In the particular case $N=1$, the multi-component extended Harry Dym
equation (\ref{ehd}) has the one-phase solution%
\begin{equation*}
v_{k}=\frac{1}{4}\sum_{m=1}^{M-k+1}ms_{m+k}r^{-m-1},
\end{equation*}%
where the function $r(\theta )$ is determined by the relationship (here $%
s_{M+1}=4c^{-2}$)%
\begin{equation*}
x+ct=\theta =\overset{r}{\int }\frac{\lambda ^{M/2}\,d\lambda }{\sqrt{%
s_{M+1}+s_{M}\lambda +\dots +s_{1}\lambda ^{M}+s_{0}\lambda ^{M+1}+\kappa
^{2}\lambda ^{M+2}}};
\end{equation*}%
and the $\kappa $-deformed one-phase solution%
\begin{equation*}
v_{k}=\frac{1}{4}e^{2\kappa x}\sum_{m=1}^{M-k+1}ms_{m+k}R^{-m-1},
\end{equation*}%
where the function $R(\vartheta )$ is determined by the relationship (here $%
s_{M+1}=4c^{-2}$)%
\begin{equation*}
\frac{e^{\kappa x}-1}{\kappa }+ct=\vartheta =\overset{R}{\int }\frac{\lambda
^{M/2}\,d\lambda }{\sqrt{s_{M+1}+s_{M}\lambda +\dots +s_{1}\lambda
^{M}+s_{0}\lambda ^{M+1}}}.
\end{equation*}

\subsection{The Extended Harry Dym Equation}

Here we consider the particular case $M=1$, i.e., the extended Harry Dym
equation%
\begin{equation*}
v_{t}=-\tfrac{1}{2}(\partial _{x}^{3}-\kappa ^{2}\partial _{x})v^{-1/2}.
\end{equation*}%
Its $N$-phase solutions are determined by\footnote{%
The simplest case $N=1$ is presented in the Introduction.}%
\begin{equation*}
v=\frac{s_{2N}}{4}\left( \prod_{n=1}^{N}r^{n}\right) ^{-2},
\end{equation*}%
where%
\begin{equation*}
x=\underset{m=1}{\overset{N}{\sum }}\,\overset{r^{m}}{\int }\frac{\lambda
^{N-1/2}d\lambda }{\sqrt{S(\lambda )}},\quad
t=\frac{\sqrt{s_{2N}}}{2}%
\underset{m=1}{\overset{N}{\sum }}\,\overset{r^{m}}{\int }\frac{\lambda
^{1/2}d\lambda }{\sqrt{S(\lambda )}},
\end{equation*}%
\begin{equation*}
0=\underset{m=1}{\overset{N}{\sum }}\,\overset{r^{m}}{\int }\frac{\lambda
^{k+1/2}d\lambda }{\sqrt{S(\lambda )}},\quad k=1,\ldots ,N-2,
\end{equation*}%
and%
\begin{equation*}
S(\lambda )=s_{2N}+s_{2N-1}\lambda +\cdots +s_{1}\lambda
^{2N-1}+s_{0}\lambda ^{2N}+\kappa ^{2}\lambda ^{2N+1}.
\end{equation*}%
A new class of solutions ($\kappa $-deformed $N$-phase solutions) is
determined by%
\begin{equation*}
v=\frac{s_{2N}}{4}e^{2\kappa x}\left( \prod_{n=1}^{N}R^{n}\right) ^{-2},
\end{equation*}%
where%
\begin{equation*}
\frac{e^{\kappa x}-1}{\kappa }=\underset{m=1}{\overset{N}{\sum }}\,\overset{%
R^{m}}{\int }\frac{\lambda ^{N-1/2}d\lambda }{\sqrt{P(\lambda )}},\quad%
t=\frac{\sqrt{s_{2N}}}{2}\underset{m=1}{\overset{N}{\sum }}\,\overset{R^{m}}{%
\int }\frac{\lambda ^{1/2}d\lambda }{\sqrt{P(\lambda )}},
\end{equation*}%
\begin{equation*}
0=\underset{m=1}{\overset{N}{\sum }}\,\overset{R^{m}}{\int }\frac{\lambda
^{k+1/2}d\lambda }{\sqrt{P(\lambda )}},\quad k=1,\ldots ,N-2,
\end{equation*}%
and%
\begin{equation*}
P(\lambda )=s_{2N}+s_{2N-1}\lambda +\cdots +s_{1}\lambda
^{2N-1}+s_{0}\lambda ^{2N}.
\end{equation*}

The case $N=1$ for the extended Harry Dym equation was considered in the
Introduction.

\section{Conclusion}

\label{sec:final}

Using the multi-component extended Harry Dym equation as an example, we studied 
integrable systems connected with their high-frequency limits ($\kappa =0$) by an 
invertible point transformation and obtained a new class of their solutions.
Applying transformation (\ref{zet}), the multi-phase solutions of the high-frequency
limits could be recalculated into a new kind of solutions for the original systems. 
As a future perspective, one can apply the $\kappa $-transformation to, e.g., 
multi-peakon solutions of the Extended Harry Dym equation to obtain a new class 
of solutions for its high-frequency limit, well-known as the Hunter--Saxton 
equation (see again \cite{Dai}), etc.

\section*{Acknowledgements}

MM gratefully acknowledges the support from GA\v{C}R under project
P201/12/G028. MVP's work was partially supported by the grant of Presidium
of RAS \textquotedblleft Fundamental Problems of Nonlinear
Dynamics\textquotedblright\ and by the RFBR grant 14-01-00012. MVP thanks
V.E. Adler, L.V. Bogdanov, E.V. Ferapontov, V.G. Marikhin, A.I. Zenchuk for
important discussions.


\end{document}